\documentstyle[12pt]{article}
\def\be{\begin{equation}}
\def\ee{\end{equation}}
\def\bear{\begin{eqnarray}}
\def\eear{\end{eqnarray}}
\def\ra{\rightarrow}
\def\cp{$CP~$}

\def\lessapprox{\stackrel{_<}{_\sim} }
\begin{document}
\begin{flushright}
PRL-TH-94/35\\
November 1994\\
hep-ph/9411398
\end{flushright}
\begin{center}
{ \Large \bf CP violation at colliders}\footnote{Based on
invited talk presented at the {\it Workshop on High Energy Particle
Physics III}, Madras, January 10-22, 1994.}
\vskip .5cm
Saurabh D. Rindani
\vskip .2cm
{\it Theory Group, Physical Research Laboratory}\\
{\it Navrangpura, Ahmedabad 380009, India}
\vskip .5cm
{\bf Abstract}
\end{center}
\vskip .2cm

The prospects of experimental detection of $CP$ violation at $e^+e^-$
and $pp/p\overline{p}$ colliders are reviewed. After a
general discussion on the quantities which can measure $CP$ violation
and on the implications of the $CPT$ theorem, various possibilities of
measuring $CP$ violation arising outside the standard model are taken up.
$CP$ violation in leptonic processes, especially polarization effects in
$e^+e^-\rightarrow l^+l^-$ are discussed next. $CP$ violation in
$t\overline{t}$ and $W^+W^-$ production and decay is also described.

\vskip 1cm
\begin{center}
{\large \bf 1. Introduction}
\end{center}
\vskip .2cm
\noindent {\bf {\boldmath 1.1 $CP$ violation in the standard model}}
\vskip .2cm

$CP$ violation in the standard model (SM), as is well known, arises due to
complex Yukawa couplings, and finally shows up through quark mixing in
the Cabibbo-Kobayashi-Maskawa matrix as a single phase. The
reparametrization invariant quantity
\be
J=\sin^2 \theta_1\sin \theta_2
\sin \theta_3\cos \theta_1\cos \theta_2\cos \theta_3\sin\delta
\ee
is a measure of $CP$ violation in SM, and though small, produces measurable
effects in  the $K$ meson system, and hopefully the effects in the $b$-quark
states will be measurable. The effects in other sectors (as for example the
neutron electric dipole moment) and in high-energy processes is
generally predicted to be too small to be observed.

In the leptonic sector of SM the prospects of $CP$ violation being observable
 are worse, since there is no analogue of the CKM phase in the absence of
 neutrino masses. $CP$ violation in leptonic systems has to feed in from the
hadronic sector through loops. For example, electric dipole moments (EDM) of
charged leptons are induced due to the $W$ EDM, and are found to vanish up to
three-loop order. The electron EDM is then estimated to be around 10$^{-41}$
$e$ cm. \cite{booth}

Thus, any new observable $CP$ violation would be a signal of
non-standard physics. It might be mentioned that perhaps there
is already a hint towards non-standard  $CP$ violation in
current ideas on electroweak baryogenesis.
\vskip .5cm
\noindent{\bf {\boldmath 1.2 Scenarios for $CP$ violation beyond SM}}
\vskip .2 cm

$CP$ violation beyond SM can arise due to almost any extra Yukawa couplings
which can be complex, and possibly also due to new Higgs self-interactions
and complex scalar vacuum expectation values. Thus, introduction of extra
 fermions or scalars could give rise to new sources of $CP$ violation.

Retaining the gauge group to be $SU(2)_L\times U(1)$, $CP$ violation can arise
due to extra fermion or Higgs doublets or singlets. Since the SM measure
of $CP$ violation $J$ (eq.(1)) is small owing to the small mixing angles
among quark generations as experimentally observed, larger $CP$ violating
 effects would arise if there are extra generations of quarks, whose mixing
angles may be less constrained.  If there are new fermions
(quarks or leptons) in exotic representations
(left-handed singlets and/or right-handed doublets of $SU(2)_L$)
there are further complex flavour-changing couplings to $Z$
which violate $CP$.

Supersymmetry requires the addition of extra scalars and fermions, whose
 couplings violate $CP$. In left-right symmetric models, again,
there further sources of $CP$ violation.
\vskip .5cm
\noindent {\bf 1.3 Use of effective Lagrangians for model-independent analysis}
\vskip .2cm

There is a large variety of sources of $CP$ violation beyond SM, and rather
than discuss predictions of each model for each observable quantity, it is more
economical to analyze $CP$-violating quantities in terms of the parameters
of an effective Lagrangian. Examples of $CP$-violating terms in an  effective
Lagrangian with which we will be concerned here are given below:
\bear
{\cal L}_{eff}&=&-\frac{i}{2}\sum_i d_{\psi_i}\overline{\psi}_i
		\sigma^{\mu\nu}\gamma_5\psi_i F_{\mu\nu}
		-\frac{i}{2}\sum_i \widetilde{d}_{\psi_i}\overline{\psi}_i
                \sigma^{\mu\nu}\gamma_5\psi_i (\partial_{\mu}Z_{\nu}-
		\partial_{\nu}Z_{\mu})\nonumber \\
	&& + \sum_{V=A,Z} i g_V \left[\widetilde{\kappa}^VW^{\dagger}_{\mu}
		W_{\nu}\widetilde{V}^{\mu\nu} +
                 \frac{\widetilde{\lambda}^V}{m^2_W}
W_{\lambda\mu}^{\dagger}W^{\mu}_{\nu}\widetilde{V}^{\nu\lambda}\right.
	\nonumber \\
&&	 \left. + g^V_4 W^{\dagger}_{\mu}
                W_{\nu}(\partial ^{\mu} V^{\nu}+\partial ^{\nu} V^{\mu})
		\right] .
\eear
$(\widetilde{V}^{\mu\nu}\equiv \frac{1}{2}\epsilon
^{\mu\nu\alpha\beta}V_{\alpha\beta};\, V_{\mu
\nu}=\partial_{\mu} V_{\nu}-\partial_{\nu} V_{\mu};
\,W_{\mu\nu}=\partial_{\mu} W_{\nu}-
\partial_{\nu} W_{\mu})$.
These terms are of dimension $\leq 6$. In eq.(2), $\psi_i$
refers to various fermionic
fields (quarks and leptons), whose electric and ``weak" dipole
moment is given by
$d_{\psi_i}$ and $\widetilde{d}_{\psi_i}$, respectively. It should
be noted that all the
 parameters are in reality scale dependent ``form factors", and can be complex.
\vskip 1cm
\noindent {\large \bf 2. Some general considerations}
\vskip .5 cm
\noindent {\bf {\boldmath 2.1 Observable quantities which measure
$CP$ violation}}
\vskip .2cm

There are basically two types of observables which can be used
to characterize $CP$
violation: asymmetries and correlations.

An example of an asymmetry is the partial-width asymmetry for
decay of particles
$i$ and $\overline{i}$ which are $CP$ conjugates of each other:
\be
A= \frac{\Gamma (i\ra f) -\Gamma (\overline{i} \ra \overline{f})}
	{\Gamma (i\ra f) +\Gamma (\overline{i} \ra \overline{f})}.
\ee
If $CP$ is a symmetry of the theory, $A=0$. Non-vanishing $A$
implies violation of  $CP$. $A$ is a convenient parameter
because it is dimensionless and lies between $-1$ and 1.
In particular, if $i$ is an eigenstate of $CP$,
$\overline{i}=i$, and $A$
in (3) simplifies and measures the fractional difference in the
decay rates of $i$ to two $CP$-conjugate states $f$ and
$\overline{f}$. As we shall see later, the $CPT$ theorem implies that
$A$ is zero even if $CP$ is violated unless the amplitude has an
absorptive part which can arise because of final-state
interactions or loop effects in perturbation theory.

Another type of asymmetry is an asymmetry in a final-state
variable like energy or angle. It is defined in general (for
$i=\overline{i}$) as
\be
A=\frac{\sum_{f\in S} d\sigma (i\ra f) - \sum_{f\in\overline{S}}
d\sigma (i\ra f)}
{\sum_{f\in S} d\sigma (i\ra f) + \sum_{f\in\overline{S}}
d\sigma (i\ra f)}.
\ee
Here $S$ and $\overline{S}$ are sets of states with
$CP$-conjugate kinematic ranges, and $f$ is a final state
assumed to have particles conjugate to one another in pairs. An
example is energy asymmetry between the energies $E_+$ and $E_-$
of $CP$-conjugate particles in $f$:
\be
A=\frac{\int_{E_+>E_-} d\sigma (i\ra f) - \int_{E_+<E_-}
d\sigma (i\ra f)}
{\int_{E_+,E_-} d\sigma (i\ra f) }.
\ee

The other category of quantities consists of $CP$-odd
correlations which are expectations values of $CP$-odd operators
in a process with initial as well final states described by
$CP$-even density matrices. Thus for an observable ${\cal
O}\left( \left\{ {\bf p}_{A_i}, {\bf s}_{A_i}\right\} \right)$
which is a function of momenta ${\bf p}_{A_i}$ and spins ${\bf
s}_{A_i}$ of particles $A_i$, and which transforms under $CP$ as
\be
{\cal
O}\left( \left\{ {\bf p}_{A_i}, {\bf s}_{A_i}\right\} \right)
\ra {\cal
O}\left( \left\{- {\bf p}_{A_i}, {\bf s}_{A_i}\right\} \right)
= - {\cal
O}\left( \left\{ {\bf p}_{A_i}, {\bf s}_{A_i}\right\} \right) ,
\ee
the $CP$-odd correlation is
\be
\langle {\cal O}\rangle = \frac{\int d\sigma {\cal
O}\left( \left\{ {\bf p}_{A_i}, {\bf s}_{A_i}\right\} \right) }
{\int d\sigma }.
\ee
A non-zero value of such a correlation signals $CP$ violation.

It may be noted that an asymmetry in the variable ${\cal O}$may
be described as a special case of a correlation of $\epsilon
({\cal O})$, where $\epsilon$ is the antisymmetric step function:
\be
\langle \epsilon ({\cal O})\rangle = \frac{\int_{{\cal O}>0} d\sigma
-\int_{{\cal O}<0} d\sigma
}
{\int_{{\cal O}} d\sigma  }.
\ee
\vskip .5cm
\noindent {\bf 2.2 Statistical significance}
\vskip .2cm

Whether or not a measured asymmetry or correlation can really
signal $CP$ violation naturally depends on  its statistical
significance decided by the statistical fluctuation expected in
the event sample.

For a rate asymmetry $A$, the number of asymmetric events
$\Delta N$ is
\be
\Delta N = A N,
\ee
where $N$ is the total number of events in the channel
considered. The statistical fluctuation in these $N$ events is
$\sqrt{N}$. Hence for discrimination of the signal, at the one
standard deviation level, we require
\be
\Delta N > \sqrt{N},
\ee
or
\be
A > \frac{1}{\sqrt{N}}.
\ee
Thus, it would be possible to measure an asymmetry if its
predicted value is larger than $1/\sqrt{N}$. To put it
differently, the number of events should be larger than $1/A^2$.

In the case of a $CP$-odd correlation $\langle {\cal O}\rangle$,
$CP$-invariant interactions can give individual events with
${\cal O}\neq 0$, averaging out to zero. Thus the mean square
deviation
\be
\Delta{\cal O}=\sqrt{\langle {\cal O}^2\rangle - \langle {\cal
O}\rangle ^2}
\ee
is a measure of the background coming from $CP$-invariant
interactions. For $N$ events in the channel, the $CP$-even
events give rise to a fluctuation $\Delta {\cal O}/\sqrt{N}$.
The signal $\langle {\cal O}\rangle$ should be larger than this
to be measurable at the one standard deviation level:
\be
\langle {\cal O}\rangle > \frac{1}{\sqrt{N}}\sqrt{\langle
{\cal O}^2\rangle - \langle {\cal O}\rangle ^2}.
\ee

There is a further experimental requirement for measuring $CP$
violation. All experimental cuts must respect $CP$ invariance.
If not, they would introduce artificial asymmetries, diluting or
obliterating the genuine signal of $CP$ violation.
\vskip .5 cm

\noindent {\bf {\boldmath 2.3 $CPT$ theorem and all that}}
\vskip .2cm

Since a combined $CPT$ transformation is a good symmetry
according to the $CPT$ theorem, $CP$ invariance (or violation)
implies $T$ invariance (or violation), and {\it vice versa}.
However, it should be borne in  mind that observation of a
$T$-odd asymmetry or correlation is not necessarily an
indication of $CP$ (or even $T$) violation. The reason for this
is the anti-unitary nature of the time-reversal operator in
quantum mechanics. As a consequence of this, a $T$ operation not
only reverses spin and three-momenta of all particles, but also
interchanges initial and final states. This last interchange is
difficult to meet with in practice, and one usually has a
situation where only momenta and spins are reversed, with the
initial and final states kept as such. In that case, non-zero
$T$-odd observables do not necessarily signal genuine $T$
violation.

There is, however, a case when $T$-odd observables imply $T$
violation, and that is when final-state interactions and loop
effects can be neglected. In that case the transition operator
${\cal T}$ obeys ${\cal T}={\cal T}^\dagger$, since the
right-hand side in the unitarity relation
\be
{\cal T}-{\cal T}^\dagger =i{\cal T}^\dagger {\cal T}
\ee
can be neglected. Then
\be
\langle f \vert {\cal T}\vert i \rangle \approx \langle f \vert
{\cal T}^\dagger \vert i \rangle = \langle i \vert {\cal T}\vert
f \rangle ^* .
\ee
Now if $T$ invariance holds, then
\be
\langle f \vert {\cal T}\vert i \rangle = \langle i_T \vert
{\cal T}\vert f_T \rangle ,
\ee
where $\vert i_T\rangle,\;\vert f_T\rangle$ represent
states with all momenta and spins inverted in sign. Combining
eq.(16) with (15) for time-reversed states, we have
\be
\vert\langle f \vert {\cal T}\vert i \rangle\vert=\vert \langle
f_T \vert {\cal T}\vert i_T \rangle\vert .
\ee
In this case, if a $T$-odd observable is non-zero, it implies
$T$ violation. Thus, $T$ invariance (and $CP$ invariance through
the $CPT$ theorem) implies equality of amplitudes with all
momenta and spins reversed if and only if final-state
interaction (absorptive part for the amplitude) vanishes.

Put differently, this means that final-state interactions can
mimic $T$ violation, but not genuine $CP$ vio\-lation. One should
therefore use, as far as possible, $CP$-odd observables to test
$CP$ invariance, not $T$-odd observables (unless they are also
$CP$ odd).

For a genuine $CP$-odd quantity, there are two possibilities, \\
A. it is $T$ odd, and therefore $CPT$ even, or \\
B. it is $T$ even, and therefore $CPT$ odd.\\
In case B, there is no violation of the $CPT$ theorem provided
the amplitude has an absorptive part. (This is again due to the
fact the $CPT$ operator is antiunitary, and interchanges initial
and final states). Thus non-vanishing of $CPT$-odd operators
necessarily requires an absorptive part of the amplitude to exist.

The absorptive part of $CP$-odd $CPT$-odd quantities in
perturbation  theory usually comes from loop contributions where
the intermediate state can be on shell. An interesting way of
realizing this possibility in the case of an intermediate state
of an unstable particle is through the Breit-Wigner form of its
propagator. In this case the absorptive part is proportional to
its width. This trick has been used in the case of the top, $Z$
and Higgs propagators \cite{Hoogeveen,width,Pilaftt}. One must
however be careful
to subtract out the part of the width corresponding to decay
into the initial or final state for consistency with the $CPT$ theorem
\cite{Soares}. It has also been pointed out
recently \cite{Arens} that off-diagonal contributions to
the self-energy of the virtual particles are also needed for
consistency with the $CPT$ theorem.
\vskip .5in
\begin{center}
{\large \bf \boldmath 3. $CP$ violation in the leptonic sector}
\end{center}
\vskip .5cm
\noindent{\bf {\boldmath 3.1 Scenarios for leptonic $CP$ violation}}
\vskip .2cm

In the standard model, no right-handed neutrinos are introduced.
As a result, there is no mass matrix to diagonalize for the
neutrinos. Hence the CKM matrix is the unit matrix, and no
$CP$-violating phases can arise. However, in extensions of the
standard model, \cp violation can arise either because of the
presence of neutrino masses or because of extra leptons
introduced (even though neutrinos may be massless), or both.

\vskip .2cm
\noindent {\bf A. Massive neutrinos}. Neutrinos can have Dirac or
Majorana masses. $CP$ violation in the Dirac case is exactly
analogous to that in the quark sector of the standard model. In
case of Majorana masses, the freedom of phase redefinition of
the neutral lepton fields is reduced because Majorana mass terms
are not invariant under phase transformations. As a result there
are more $CP$-violating phases in the CKM matrix than the
corresponding Dirac case. It is thus possible to have \cp
violation with even two generations of Majorana neutrinos.

 \vskip .2 cm
\noindent {\bf B. Massless neutrinos with exotic leptons}. It is
possible to have $CP$ violation because of either charged or
neutral leptons in exotic representations of $SU(2)\times U(1)$.
The leptons then have flavour-violating couplings to $Z$ or
Higgses, which can be complex and hence $CP$ violating.

\vskip .2cm
We consider below some leptonic $CP$-violating processes at
high-energy colliders which make use of the above mechanisms of
$CP$ violation. The importance of leptonic processes stems from
the fact that they are relatively clean from the experimental
point of view.

\vskip .5cm
\noindent {\bf \boldmath 3.2 Leptonic flavour violating $Z$ decays}
\nopagebreak
\vskip .2cm
Leptons can have flavour-violating couplings to $Z$ giving rise
to flavour violating $Z$ decays into charged leptons either at
tree level or at one-loop level:
\be
Z\ra l_i\overline{l_j}\;\;(i\neq j).
\ee
The corresponding $CP$-violating asymmetry is
\be
A= \frac{\Gamma (Z\ra l_i\overline{l_j})-\Gamma (Z\ra \overline{l_i}l_j)}
{\Gamma (Z\ra l_i\overline{l_j})+\Gamma (Z\ra \overline{l_i}l_j)}.
\ee
This is $T$ even and therefore $CPT$ odd. It therefore needs an
absorptive part to be present.

Flavour-violating tree-level couplings of charged leptons to $Z$
arise in models with exotic charged leptons transforming as
either left-handed singlets and/or right-handed doublets. In
such a case the Glashow-Weinberg condition for flavour-diagonal
couplings is not satisfied, and (18) occurs at tree level. For
$A$ of (19) to be non-zero, one-loop correction to (18) is also
needed, and only the absorptive part of that amplitude
contributes. The asymmetry is then ${\cal O}(\alpha
)$\cite{Choudhury}.

On the other hand, models with exotic neutral leptons have
flavour-violating couplings of neutral leptons at the tree-level
giving rise to (18) at the loop level \cite{Rius}. The
absorptive part now comes from one of these loop diagrams. $A$
is now ${\cal O}(1)$. However, unlike in the previous case, the
rate of the flavour-violating process (18) is ${\cal O}(\alpha
^3)$. Thus, the minimum total number $N_Z$ of $Z$ events for an
observable asymmetry is in both cases ${\cal O}(1/\alpha ^3)$.
However, constraints on leptonic mixing angles and masses of
exotic leptons make this process too rare to observe at LEP.

\vskip .5cm
\noindent {\bf {\boldmath 3.3 $CP$ violation in $e^+e^-\ra \gamma ^*,Z^* \ra
l^+l^-$} }
\vskip .2cm

In case of $CP$ violation in $e^+e^-\ra \gamma ,Z \ra l^+l^-$
there are two general results \cite{BernLow}: \\
\noindent (i) No $CP$ violation can be seen without measuring
initial or final spins. This follows basically because no $CP$-odd
observable can be constructed without spins. \\
\noindent (ii) The only $CP$-violating couplings for the
on-shell process are the dipole moment type couplings of $e$ or
$l$ (electric or ``weak" dipole moments). Since
there are strong experimental limits on the electron and muon
electric dipole
moments ($d_e \lessapprox 10^{-27} e$ cm,
$d_{\mu}\lessapprox 10^{-19} e$ cm), $\tau$ may be a good
candidate for $l$. In fact, the weak moment of $\tau$ has been
constrained using $\tau$ polarization in this reaction (see below).

Since it is clear from (i) that either initial or final
spins have to be observed to detect $CP$ violation, we consider
below both these cases for $e^+e^-\ra Z \ra {\tau}^+{\tau}^-$.

\vskip .25cm
\noindent {\bf {\boldmath 3.3.1 $CP$ tests using $\tau$ polarization in
$e^+e^- \ra \tau ^+\tau^-$}}
\vskip .2cm

This possibility has been discussed by several authors
[2,10--12]. For the process
\be
e^-(p_-)+e^+(p_+)\ra \tau^-(k_-,s_-)+\tau^+(k_+,s_+),
\ee
possible $CP$-odd quantities that can be constructed out of
the momenta and spins in the centre-of-mass (cm) frame are
$({\bf p}_- - {\bf p}_+)\cdot ({\bf s}_- - {\bf s}_+)$ and
$({\bf p}_- - {\bf p}_+)\cdot {\bf s}_-\times {\bf s}_+$.

To measure these quantities, one must, of course, be able to
measure ${\bf s}_{\pm}$. This can be done by looking at decay
distributions of $\tau _{\pm}$. In the rest frame of $\tau$, the
angular distribution of an observed decay particle is
\be
\frac{d\Gamma}{d\Omega ^*} = \frac{1}{4\pi }\left( 1+\alpha {\bf
s}\cdot \widehat{\bf q}^*\right) ,
\ee
where $\widehat{\bf q}^*$ is the unit vector along the momentum of
the  observed particle, and $\alpha$ is a constant called the
analyzing power of the channel. For example,
\begin{center}
$
\begin{array}{ccccc}
\alpha &= & \mp 1 & {\rm for}& \tau ^{\pm}\ra \pi^{\pm}\nu
_{\tau}, \\
       &= & \pm 1/3 &{\rm for}&  \tau ^{\pm}\ra l^{\pm}\nu _l\nu
_{\tau},
 \end{array} $ \\
\end{center}
as deduced from the theory of weak $\tau$ decay. Using (21),
spin correlations can be translated to momentum correlations.

In terms of the observed momenta, possible $CP$-odd variables are
$\widehat{\bf p}\cdot ({\bf q}_+\times {\bf q}_-)$ ($CPT=+1$), $\widehat{\bf
p}\cdot ({\bf q}_+ + {\bf q}_-)$ ($CPT= -1$),
$\widehat{\bf p}\cdot ({\bf q}_+\times {\bf q}_-)\, \widehat{\bf
p}\cdot ({\bf q}_+ - {\bf q}_- )$  ($CPT=+1$), $\widehat{\bf
p}\cdot ({\bf q}_+ + {\bf q}_- ) \widehat{\bf
p}\cdot ({\bf q}_+ - {\bf q}_- )$ ($CPT= -1$).
Expectation values of the last two were suggested by Bernreuther
{\it et al}. \cite{BernZPC,BernPRD} for measuring respectively the real and
imaginary parts of the $\tau$ weak dipole form factor
$\widetilde{d}_{\tau} (m^2_Z)$. The suggestion has been carried out at
LEP for Re$\,\widetilde{d}_{\tau} (m^2_Z)$ by the OPAL \cite{OPAL} and ALEPH
\cite{ALEPH} groups. OPAL looked at inclusive leptonic and hadronic
decays of $\tau$, whereas ALEPH analyzed all channels
exclusively. The results obtained are the 95\% confidence-level upper
limits Re$\,\widetilde{d}_{\tau} (m^2_Z) <7.0\times 10^{-17} e$ cm (OPAL
\cite{OPAL}) and Re$\,\widetilde{d}_{\tau} (m^2_Z) <3.7\times 10^{-17} e$ cm
(ALEPH \cite{ALEPH}).

The theoretical prediction for the 1 s.d. limit obtainable in
the measurement of Im$\,\widetilde{d}_{\tau} (m^2_Z)$ is $10^{-16}$
using the correlation $\langle\widehat{\bf
p}\cdot ({\bf q}_+ + {\bf q}_- ) \widehat{\bf
p}\cdot ({\bf q}_+ - {\bf q}_- )\rangle$ and a sample of $10^7$
$Z$'s \cite{BernPRD}.
\vskip .25cm
\noindent{\bf 3.3.2 Longitudinal beam polarization}
\vskip .2cm

The Stanford Linear Collider (SLC), operating presently at the
$Z$ resonance, has an $e^-$ polarization of about 62\%, and is
likely to reach 75\% in the future. The present sample collected
is of 50,000 $Z$'s, and the hope is to reach 5$\times10^5$, or
even $10^6$ $Z$'s.

Can this longitudinal $e^-$ polarization help in measuring the
$\tau$ weak dipole moment? The answer is ``yes" \cite{Banantha}. In fact,
as we shall see, SLC can do better than LEP so far as
Im$\widetilde{d}_{\tau}$ is concerned.

The essential point is that the vector polarization of $Z$ gets
enhanced in the presence of $e^+e^-$ longitudinal polarization.
For vanishing beam polarization, $P_{e^-} = P_{e^+} =0$, the $Z$
vector polarization is
\be
P_Z^{(0)} = \frac{2 g_{Ve}g_{Ae}}{g^2_{Ae}+g^2_{Ve}} \approx 0.16,
\ee
where $g_{Ve}$, $g_{Ae}$ are vector and axial-vector couplings
of $e^+e^-$ to $Z$. For non-zero polarization,
\be
P_Z = \frac{P_Z^{(0)}-P_{e^+e^-}}{1-P_Z^{(0)}P_{e^+e^-}},
\ee
where
\be
P_{e^+e^-} = \frac{P_{e^-}-P_{e^+}}{1-P_{e^-}p_{e^+}}.
\ee
Thus, $P_Z\approx 0.71$ for $P_{e^-}=-.62$ and $P_{e^+}=0$,
which is an enhancement by a factor of about 4.5.

It is therefore profitable to look for $CP$-odd observables
involving the $Z$ spin ${\bf s}_Z=P_Z\widehat{\bf p}$, where
$\widehat{\bf p}$ is the unit vector along ${\bf p}_+=-{\bf p}_-$ in
the c.m. frame. Examples of such observables are $\widehat{\bf
p}\cdot ({\bf q}_+\times {\bf q}_-)$ and $\widehat{\bf
p}\cdot ({\bf q}_+ + {\bf q}_-)$. While both are $CP$ odd, the
former is $CPT$ even and the latter is $CPT$ odd.

The above is not entirely correct in principle. $CP$-odd
correlations give a measure of underlying $CP$ violation only if
the initial state is $CP$ even. Otherwise there may be
contributions to correlations which arise from $CP$-invariant
interactions due to the $CP$-odd part of the initial state. In
the case when only the electron beam is polarized, the initial
state is not $CP$ even. In practice, however, this $CP$-even
background is small because for $m_e \ra 0$, only the $CP$-even
helicity combinations $e_L^-e_R^+$ and $e_R^-e_L^+$ survive,
making the corrections proportional to $m_e/m_Z\approx 5\times
10^{-6}$. If one includes order $\alpha$ collinear photon
emission from the initial state, which could flip the helicity
of the $e^{\pm}$, then like-helicity $e^+e^-$ states could also
survive for vanishing electron mass\footnote{I thank Prof. L.M.
Sehgal for drawing my attention to this fact.}. However, it turns
out that this being a non-resonant effect, the corresponding
cross section at the $Z$ peak is small. It is therefore expected
that the correlations coming from $CP$-invariant SM interactions
in such a case will be negligible.

The correlations $\langle O_1\rangle \sim \langle \widehat{\bf
p}\cdot ({\bf q}_+\times{\bf q}_-)\rangle$ and $\langle
O_2\rangle \sim \langle \widehat{\bf
p}\cdot ({\bf q}_+ + {\bf q}_-)\rangle$ have been calculated
analytically for the single-pion and $\rho$ decay mode of each
$\tau$. Also calculated analytically are $\langle O_1^2\rangle$
and $\langle O_2^2\rangle$ needed for obtaining the 1 s.d. limit
on the measurability of $\widetilde{d}_{\tau}$ obtained using
eq.(13) \cite{Banantha}.

 The results for only the single-pion channel are
summarized in Tables 1a and 1b, which give, respectively for
$O_1$ and $O_2$ and for electron polarizations $P_{e^-}=0,\pm 0.62$,
the correlations in units of $\widetilde{d}_{\tau}m_Z/e$, the square
root of the variance, and the 1$\sigma$ limit on ${\rm
Re}\,\widetilde{d}_{\tau}$ and ${\rm Im}\,\widetilde{d}_{\tau}$ obtainable
with $10^6$ $Z$'s. The enhancement of $\langle O_{1,2}\rangle$ and
hence the sensitivity of $\widetilde{d}_{\tau}$ measurement with
polarization is evident from the tables.

\vskip .5cm

\begin{center}
\begin{tabular}{||c|c|c|c||}
\hline
$P_e$ &$\langle O_1\rangle\,$(GeV$^2$) for
&$\sqrt{\langle  O^2_1\rangle }$&1 s.d. limit on
Re$\,\widetilde{d}_{\tau}$\\
&Re$\,\widetilde{d}_\tau=e/m_Z$ &(GeV$^2$)
&for $10^6\, Z$'s (in $e$ cm)\\
\hline
{}~0 &~0.90 &$12.86$ &$1.5\times 10^{-16}$\\
$+0.62$ &$-2.89$ &12.86 &$4.6\times 10^{-17}$\\
$-0.62$ &$~4.01$ &12.86 &$3.3\times 10^{-17}$\\
\hline
\end{tabular}
\vskip 0.2cm
Table 1a
\end{center}
\vskip .5cm
\begin{center}
\begin{tabular}{||c|c|c|c||}
\hline
$P_e$ &$\langle O_2\rangle\,$(GeV) for
&$\sqrt{\langle  O^2_2\rangle }$&1 s.d. limit on
Im$\,\widetilde{d}_{\tau}$\\
&Im$\,\widetilde{d}_\tau=e/m_Z$ &(GeV)
&for $10^6\, Z$'s (in $e$ cm)\\
\hline
{}~0 &$-0.16$ &9.57  &$6.2\times 10^{-16}$\\
$+0.62$ &$~0.51$ &9.57 & $1.9\times 10^{-16}$\\
$-0.62$ &$-0.70$ &9.57 &$1.4\times 10^{-16}$\\
\hline
\end{tabular}
\vskip 0.2cm
Table 1b
\end{center}
\vskip .5cm
\noindent {\small Tables 1a and 1b: Correlations of $O_1$ and $O_2$
respectively, the standard deviation from SM interactions, and
the 1 s.d. limit on the real and imaginary parts of the weak
dipole moment, for different polarizations $P_e$.}

\vskip 1cm
The sensitivity can be further improved by looking at only the
$P_{e^-}$ dependent part of the cross section. This can be done by
looking at a sample corresponding to the difference in the
number of events for a polarization $P_{e^-}$ and polarization
$-P_{e^-}$. The correlations are evaluated with respect to
$d\sigma (P_{e^-})-d\sigma (-P_{e^-})$. This reduces the number
of events. However, correlations are enhanced by a larger factor
giving a net gain in the sensitivity. The result for
$P_{e^-}=.62$ and $10^6$ $Z$'s is given in Table 2.

\begin{center}
\begin{tabular}{||c|c|c|c||}
\hline
Observable&$\langle O\rangle$ for&$\sqrt{\langle  O^2\rangle
}$ &1 s.d. limit on $\vert
\widetilde{d}_\tau\vert$ \\
&$\vert
\widetilde{d}_\tau\vert =e/m_Z$&&
 for $10^6\, Z$'s (in $e$ cm)\\
\hline
$O_1$ &35.55 GeV$^2$&12.86 GeV$^2$ &$1.2\times 10^{-17}$\\
$O_2$ &$-6.22$ GeV&$~9.57$ GeV &$5.0\times 10^{-17}$\\
\hline
\end{tabular}
\vskip 0.5cm
\end{center}
\noindent {\small Table 2: Quantities as in Table 1, but for a
distribution asymmetrized between polarizations $+0.62$ and $-0.62$.}
\vskip .5cm

The 1 s.d. limit for Im$\widetilde{d}_{\tau}$ should be compared with
the LEP expectation of $10^{-16}$ $e$ cm for a larger sample of
$10^7$ $Z$'s \cite{BernPRD}.

The use of correlations for measuring the $\tau$ edm at the proposed
$\tau$-charm factories employing longitudinal polarization of electron
and positron beams has also been studied in detail in \cite{taucf}.

\vskip .25cm
\noindent {\bf 3.3.3 Transverse beam polarization}\nopagebreak
\vskip .2cm

Use of transversely polarized $e^+e^-$ beams for the study of
$CP$ violation has been studied by several people
(see for example \cite{Hoogeveen,Burgess}). For a reaction
\be
e^-(p_-,s_-)+e^+(p_+,s_+)\ra f(k_+)+\overline{f}(k_-)
\ee
in the c.m. frame, where $f$ denotes a fermion, possible $T$-odd
triple products are ${\bf s}_-\cdot ({\bf p}\times {\bf k})$,
${\bf s}_+\cdot ({\bf p}\times {\bf k})$, ${\bf p}\cdot ({\bf
s}_-\times {\bf s}_+)$, ${\bf k}\cdot ({\bf s}_-\times {\bf s}_+)$
, where ${\bf p}_+=-{\bf p}_-={\bf p}$ and ${\bf k}_+=-{\bf
k}_-={\bf k}$. Of these triple products, the last two are purely
$CP$ odd, whereas only the difference of the first two is $CP$
odd. Burgess and Robinson \cite{Burgess} have done an analysis of
$e^+e^-\ra \tau^+\tau^-, c\overline{c}, b\overline{b}$ in terms of operators
\be
{\cal L}_W =
\lambda_W\left[ \overline{L}P_RD^{\mu}ED_{\mu}\phi\right] + H.c.,
\ee
\be
{\cal L}_Y= \frac{i}{2}\lambda_Y\left[ g_1
B_{\mu\nu}\left( \overline{L}\gamma ^{\mu\nu} P_R E\right)\phi\right]
+ H.c.
\ee

Their results for $\lambda_W = (400\, {\rm GeV})^{-2}$ and
$L=4.8\times 10^5\,{\rm pb}^{-1}$ at LEP are given in Table 3,
where {\cal A} is the asymmetry for ${\bf s}_-\cdot ({\bf
p}\times {\bf k} )$ given by
\be
{\cal A} =  \left[ \int d\sigma ( {\bf p}_i ,{\bf s}_i) - \int
d\sigma ( -{\bf p}_i ,-{\bf s}_i) \right]\,{\bf s}_-\cdot ({\bf
p}\times {\bf k} ).
\ee
Though this is an interesting effect, a theoretical estimate for
$\lambda_W$ is needed before
concluding whether the effect would be observable. Assuming a
systematic error of 0.1\%, the 2-$\sigma$ limits possible on
$\lambda_W$ are estimated to be (570 GeV)$^{-2}$ and (660
GeV)$^{-2}$ respectively for up- and down-type quarks.

\begin{center}
\begin{tabular}{||c|c|c|c||}
\hline
&$\tau^+\tau^-$&$c\overline{c}$&$b\overline{b}$\\
\hline
$N$ (SM events)$\times 10^{-8}$&$~6.5$&24&31\\
${\cal A}\times 10^{-5}$&$-6.1$&97&$-170$\\
${\cal A}/N\,\times 10^3$&0.9&4.0&5.5\\
\hline
\end{tabular}
\vskip .5cm
\end{center}
\noindent {\small Table 3: The transverse polarization asymmetry ${\cal
A}$ (defined in the text) compared with the standard model
events for $e^+e^-\ra\tau^+\tau^-,c\overline{c},b\overline{b}$.}
\vskip .6cm
\newpage
\begin{center}
{\large \bf \boldmath 4. $CP$ violation in top pair production}
\end{center}
\vskip .2cm

Evidence for a top quark of mass of about 174 GeV at the
Tevatron has been reported by the
CDF collaboration \cite{CDF}.  Even though the data
is not conclusive, it is generally believed that the top quark
will eventually be found with a mass of a similar magnitude.
Top-antitop pairs can then be produced at large rates at
future colliders and used for various studies. In particular,
since a heavy top ($m_t>120$ GeV) decays before it hadronizes
\cite{Kuhn}, information about its polarization is preserved in
its decay products.
Schmidt and Peskin \cite{Schmidt} have suggested (elaborating on an
old suggestion of Donoghue and Valencia \cite{Donoghue}) looking for
the asymmetry between $t_L\overline{t}_L$ and $t_R\overline{t}_R$ as a
signal for $CP$ violation (see also \cite{Kane}). Note that this
is possible only for a
heavy particle like the top quark because for a light particle,
the dominant helicity combination would be $t_L\overline{t}_R$ or
$t_R\overline{t}_L$, each being self-conjugate.

The asymmetry $N(t_L\overline{t}_L)-N(t_R\overline{t}_R)$ can be probed
through the energy spectra of prompt leptons from $t \ra
Wb;\,W\ra l\nu$. This is understood as follows.

For a heavy top, the dominant $W$ helicity in $t\ra Wb$ is 0.
Now, due to $V-A$ interaction, $b$ is produced with left-handed
helicity (neglecting the $b$ mass). Hence in the $t$ rest frame,
$W^+$ momentum is dominantly along the $t$ spin direction. It
follows that $l^+$ in $W^+$ decay is produced preferentially in
the direction of the $t$ spin. In fact, the distribution is
$1+\cos \psi$, where $\psi$ is the angle between the
$l^+$ momentum direction and the $t$ spin direction. In going to
the laboratory frame, the $t$ gets boosted. Thus $l^+$ from
$t_R$ is more energetic than $l^+$ from $t_L$, and $l^-$ from
$\overline{t}_L$ has more energy than $l^-$ from
$\overline{t}_R$. Therefore, in the decay of
$t_L\overline{t}_L$, $l^-$ from $\overline{t}_L$ has higher
energy than $l^+$ from $t_L$, and the reverse is true for
$t_R\overline{t}_R$. Thus the energy asymmetry of leptons
measures $N(t_L\overline{t}_L)-N(t_R\overline{t}_R)$.

Schmidt and Peskin \cite{Schmidt} looked at this asymmetry in hadron collisions
in a $CP$-violating multi-Higgs model where the $CP$ violation
is described by the Lagrangian terms
\be
\delta{\cal L} = - \frac{m_t}{v} \phi \overline{t} \left[ AP_L + A^*
P_R \right] t,
\ee
where only the effect of the dominant lightest Higgs field
$\phi$ is kept. $v$ is the SM Higgs vacuum expectation value,
and $A$ is a complex combination of mixing angles and phases.

Since $N(t_L\overline{t}_L)-N(t_R\overline{t}_R)$ is $CP$ odd and $T$
even, the $CPT$ theorem requires the existence of an absorptive
part for it to be non-zero. Looking at the tree and one-loop
diagrams for $q\overline{q} \ra t\overline{t}$, they conclude that
\be
{\cal A} = \frac{N(t_L\overline{t}_L) -
N(t_R\overline{t}_R)}{N(t_L\overline{t}_L) +
N(t_R\overline{t}_R)} \approx 10^{-3}
\ee
for $m_t= 150$ GeV, $m_{\phi}=100\! -\! 400$ GeV and
Im$[A^2]=\sqrt{2}$. Thus the asymmetry ${\cal A}$ would be
observable with $10^8$ $t\overline{t}$ a year.

For $pp$ collisions, since the $pp$ state is not a $CP$
eigenstate, there is also a $CP$-invariant contribution
present,  but this is shown to be small \cite{Schmidt}.

As in the case of $pp$ or $p\overline{p}$ collisions, lepton energy
asymmetry can be used to measure $CP$ violation in $e^+e^- \ra
t\overline{t}$ \cite{Changtt}. The authors of ref.\cite{Changtt}
also consider another asymmetry, viz., the up-down asymmetry of
the charged leptons with respect to the $t\overline{t}$
production plane in the laboratory frame. It is
also possible to construct a ``left-right" asymmetry of leptons
with respect to a plane perpendicular to the $t\overline{t}$
production plane, but containing the $t\overline{t}$ momentum
direction \cite{Poulose}. Certain combinations of up-down and
left-right symmetry with
forward-backward asymmetry can also be considered. All these
probe different
combinations of $CP$-violating couplings \cite{Poulose}.

The result of ref.\cite{Changtt} is that asymmetry is at the few per cent
level for the top-quark electric and weak dipole moments $d_t,
\widetilde{d}_t\sim e/m_t$. For $\sqrt{s}=300$ GeV and $m_t=120$
GeV, $\sigma \approx 1400$ fb, which corresponds to 20,000
prompt leptons a year for a luminosity of $10^{33}$
cm$^{-2}$s$^{-1}$. In such a case, $d_t,\widetilde{d}_t$ can be
determined to a few percent level. This should be compared with
the prediction of $10^{-2}-10^{-3}$ from the Higgs model for
$CP$ violation.

Apart from the above asymmetries, $CP$-odd correlations could
provide a measure of $CP$ violation in $t\overline{t}$ production
\cite{Berntprod}. Certain correlations are more sensitive to $CP$ violation
in $t$ decay, rather than production \cite{Berntprod}.

Another process which has been suggested is $e^+e^-\ra\-
t\overline{t}\nu\overline{\nu}$ through $W^+W^-\newline\ra
t\overline{t}$ \cite{Pilaftt}. Two diagrams contribute to
$W^+W^-\ra
t\overline{t}$, one corresponding to $t$-channel $b$ exchange and the
other with an $s$-channel heavy Higgs $\phi$, which can be on
shell for $m_{\phi}>2m_t$. Then, the absorptive
part needed for a $CP$-odd $CPT$-odd asymmetry is provided by
the width of the Higgs. A sizable asymmetry can be obtained thus
\cite{Pilaftt}.

At linear colliders, there is a possibility of producing electron beams with
longitudinal polarization. This may be exploited to enhance the sensitivity of
measurememnt of the top dipole moments as well as to measure the electric and
weak dipole moments independently \cite{Poulose,Cuypers}

\vskip .6cm

\begin{center}
{\large \bf \boldmath 5. $CP$ violation in  other processes}
\end{center}

\vskip .2cm

The process $e^+e^-\ra W^+W^-$, which will be studied in the
near future at LEP200, will be the first one to be able to test
the SM couplings of the electroweak gauge bosons. The process is
expected to put bounds on non-standard $\gamma$ and $Z$
couplings to $W^+W^-$. The non-standard couplings could be $CP$
violating ones, as in (2). These can be studied in a way similar
to the one used for probing $CP$
violation in $e^+e^-\ra \tau^+\tau^-$ and $e^+e^-\ra t\overline{t}$
considered earlier. As in the case of $e^+e^-\ra t\overline{t}$ with
leptonic $t,\overline{t}$ decays, energy or up-down asymmetry of
leptons may be used \cite{ChangWW,Mani}. Whereas Chang {\it et
al}. \cite{ChangWW}
consider asymmetries, Mani {\it et al}. \cite{Mani} gave estimated the
energy correlation ratio
\[
A=\frac{\langle E_-\rangle-\langle E_+\rangle }
	{\langle E_-\rangle+\langle E_+\rangle },
\]
and suggest the angular correlation ratio
\[
\delta =\frac{\langle \theta_-\rangle-\langle \theta_+\rangle }
	{\langle \theta_-\rangle+\langle \theta_+\rangle },
\]
where $E_{\pm}$ are the energies of the leptons $l^{\pm}$ produced
in the decay of $W^{\pm}$, and $\theta_{\pm}$ are the angles of
$l^{\pm}$ momenta with respect to the $e^+$ beam direction. The
general expectation is $A\approx 10^{-3}$ for $\widetilde{\kappa}$
or $\widetilde{\lambda}\;\approx 0.1$.

Some other processes considered in the literature are $t,\overline{t}$
decay asymmetries in $\phi\ra t\overline{t}$, where $\phi$ is a
heavy Higgs \cite{Htt}, decay lepton asymmetries in $e^+e^-\ra
\widetilde{\chi}\widetilde{\chi}$, where $\widetilde{\chi}$ is a neutralino
in the minimal supersymmetric standard model \cite{kizu},
decay correlations in $\gamma\gamma\ra W^+W^-$ \cite{Ma}, and
forward-backward asymmetry in $e^+e^-\ra\gamma Z$ \cite{Munich}.

\vskip .6cm

\begin{center}
{\large \bf 6. Conclusions}
\end{center}
\vskip .2cm

We have seen above the various points to be kept in mind
when selecting processes and variables for detecting and
measuring $CP$ violation, a number of processes and scenarios
of $CP$-violating signatures which could be looked for. The
above discussion is mainly aimed at arriving at an idea of the
sensitivities possible in different measurements. In general,
the results in most popular models of $CP$ violation beyond SM
indicate that $CP$ violation in the most optimistic theoretical
scenario would be measurable only with some difficulty in the
existing or presently envisaged experiments. Nevertheless, it
would be good to keep one's eyes open to these possibilities.

\end{document}